\documentclass[final,5p,times,twocolumn,sort&compress]{elsarticle}

\usepackage{amssymb}

\journal{Journal of Magnetism and Magnetic Materials}

\newcommand{\kk}{\mathbf{k}}
\newcommand{\I}{\mathrm{i}}
\newcommand{\nVH}{n_{\rm vH}}

\begin{document}

\begin{frontmatter}



\title{Correlation Effects and Non-Collinear Magnetism in the Doped Hubbard Model}


\author[Ekb,Urfu]{P.A. Igoshev}
\author[Izh]{M.A. Timirgazin}
\author[Izh]{V.F. Gilmutdinov}
\author[Izh]{A.K. Arzhnikov}
\author[Ekb]{V.Yu. Irkhin}

\address[Ekb]{Institute of Metal Physics, Russian Academy of Sciences, 620990 Ekaterinburg, Russia}
\address[Urfu]{Ural Federal University, 620002 Ekaterinburg, Russia}
\address[Izh]{Physical-Technical Institute, 426000, Kirov str. 132, Izhevsk, Russia}

\begin{abstract}
The ground--state magnetic phase diagram is investigated for the two-- and three--dimensional $t$--$t'$ Hubbard model.
We take into account commensurate ferro--, antiferromagnetic, and incommensurate (spiral) magnetic phases, as well as phase separation into magnetic phases of different types, which was often missed in previous investigations.
We trace the influence of correlation effects on the stability of both spiral and collinear magnetic order by comparing the results of employing both the generalized non-correlated mean--field (Hartree--Fock) approximation and generalized slave boson approach by Kotliar and Ruckenstein with correlation effects included.
We found that the spiral states and especially ferromagnetism are generally strongly suppressed up to non-realistic large Hubbard $U$, if the correlation effects are taken into account.
The electronic phase separation plays an important role in the formation of magnetic states and corresponding regions are wide, especially in the vicinity of half--filling.
The details of magnetic ordering for different cubic lattices are discussed.

\end{abstract}

\begin{keyword}
incommensurate magnetism \sep electronic correlation \sep band magnetism


\end{keyword}

\end{frontmatter}


\section{Introduction}
\label{sec:intro}

The investigation of the one-band Hubbard model is an actual problem already for more than a half century.
In recent decades, the two dimensional (2D) case closely related to the problem of high temperature superconductivity has been intensively studied.

The ground state of the model in the half--filled case for the bipartite lattices is a N\'eel antiferromagnetic (AFM) insulator.
The types of instability of the antiferromagnetic state in the presence of doping or the finite integral of the electron transfer between the second neighbors are still not incompletely revealed.
According to the classical work \cite{Nagaoka66}, when one charge carrier is added, the ground state on the bipartite lattice is the saturated ferromagnetic (FM) one.
This statement can be also considered as a reasonable hypothesis in the case of finite doping~\cite{Nagaoka66,Linden91,Irkhin04}.

Scenarios of possible doping--induced magnetic ordering include the phase separation (PS) of different types: to the ferromagnetic and antiferromagnetic phases~\cite{Visscher73} or the phase of the superconducting electron liquid and the N\'eel antiferromagnetic phase~\cite{Emery90}.
An alternative scenario is the formation of the spiral (incommensurate) magnetic state. It was considered within different approaches: the analysis of the momentum dependence of the generalized static magnetic susceptibility for the bare spectrum~\cite{Schultz90}, the Hartree--Fock approximation (HFA), small and moderate $U/W$ values being treated, where $U$ is the parameter of the Coulomb repulsion and $W$ is the bandwidth~\cite{Arrigoni91,Igoshev10}, and the $t-J$ model (large $U/W$ values)~\cite{Shraiman90}.

The experimental observation of antiferromagnetic or spiral magnetic structures makes the classical problem of theoretical description of different type magnetic order formation very actual.
Spiral structures are observed in both two and three dimensional compounds: 
in iron based high temperature superconductors~\cite{Bao09} in AFM chromium and fcc $\gamma$--iron.
In doped cuprates the spiral states are found as the dynamic magnetic order~\cite{HTSCrewiew}.
Besides that, considerably enhanced incommensurate magnetic fluctuations are observed in strontium ruthenates at low temperatures~\cite{Ru} (see discussion in Refs.~\cite{Igoshev10,Igoshev07,Igoshev11}.

A detailed study of the magnetic phase diagram of the 2D Hubbard model taking into account the electron transfer only between the nearest neighbors $t$ within HFA shows that the spiral magnetic states occur in a wide range of parameters, especially at moderate values $U\lesssim W$~\cite{Sarker91}.
It was shown in~\cite{Igoshev10} that inclusion of the electron transfer between second neighbors ($t'\ne0$) in the Hamiltonian changes considerably the ground state magnetic phase diagram.
In addition to the spiral structures, the formation of so called stripes was predicted~\cite{Kivelson03,Vedyaev}.
However, this conclusion is somewhat devaluated by disregarding the intersite Coulomb interaction, which can considerably increase the energy of the stripes as inhomogeneous structures.

A convenient approach to study the formation of the magnetic order with account of correlations is the slave boson approximation (SBA) by Kotliar and Ruckenstein~\cite{Kotliar86}. In the saddle point approximation, this method is qualitatively similar to the Gutzwiller approximation~\cite{Gutzwiller}.



The effect of  electron correlations on  stability of the spiral magnetic states was considered in~\cite{Fresard92} using SBA.
The phase diagram of the Hubbard model was built in the the nearest-neighbor approximation  ($t' = 0$).
Later, the generalized static magnetic susceptibility was studied within the same method~\cite{Fresard98}.
This makes it possible to determine the criterion of the instability of the paramagnetic state with respect to a second--order transition into the incommensurate magnetic state (this generalizes the  criterion obtained within the random phase approximation~\cite{Moriya} to the  strongly correlated states).
As a result, a considerable tendency to ferromagnetic ordering at hole doping and large $t'/t$ values was found.
However, the phase transitions between the magnetically ordered states cannot be studied within this approach.

Although the description of thermodynamics is a difficult problem, in the ground state the approaches discussed can be used also for three--dimensional (3D) lattices (e.g., transition metal compounds).
In the present paper we apply both non-correlated mean-field (Hartree--Fock) approximation and the slave boson approximation to 2D and 3D $t$--$t'$ Hubbard model.


\section{Model and methods}
\label{sec:method}

We consider the Hubbard model
\begin{equation}
      \label{eq:original_H}
      \mathcal{H}=\sum_{ij\sigma} t_{ij} c^\dag_{i\sigma}c^{}_{j\sigma}+U\sum_i n_{i\uparrow}n_{i\downarrow},
\end{equation}
where the matrix elements of the electronic transfer are $t_{ij} = -t$ for the nearest neighbors and $t'$  for the next--nearest neighbors, $c^\dag_{i\sigma},c^{}_{i\sigma}$ are electronic creation and annihilation operators correspondingly.

Local spin space rotation matching different site magnetization vectors  by the angle $\mathbf{QR}_i$ (where $\bf Q$ is a spiral wave vector, ${\bf R}_i$ is a site position) is applied for the consideration of magnetic spirals. This transforms spiral magnetic state into the effective ferromagnetic one, but the hopping term in the Hamiltonian becomes non--diagonal with respect to spin index. The Hartree--Fock treatment of many--particle Coulomb interaction term replaces it by the one--electron interaction term with some effective field $Un_{i\bar\sigma}$ which however mixes the contributions of singly occupied states and doubles on the equal footing ($n_{i\sigma}$ is an average electronic density at site $i$ and spin projection $\sigma$). This is not correctly physically, especially at large $U$.

A simple way to take into account correlation effects (energetically unfavorability of doubly occupied site states) is the introducing of the slave boson operators $e_i(e_i^\dag), p_{i\sigma}(p_{i\sigma}^\dag), d_i(d_i^\dag)$, describing the transitions between site states originating from the electronic system dynamics on an alternative language but {\it simultaneously} with the conventional Slater determinant based formalism  (related with one--electron operators). Conceptually this is close to the Hubbard $X$--operator formalism\cite{Hubbard2,Irkhin04}, where however the site transition $X$--operators are introduced {\it instead of} the original one--electron operators.
The extension of the original Hubbard Hamiltonian (\ref{eq:original_H}) has the form where the Coulomb interaction term is diagonal with respect to the boson operators:
\begin{equation}
      \label{eq:H_ext}
      \mathcal{H}_{\rm ext}=\sum_{ij\sigma\sigma'} t^{\sigma\sigma'}_{ij} c^\dag_{i\sigma}c^{}_{j\sigma'}
z^\dag_{i\sigma}z^{}_{j\sigma'}
+U\sum_i d^\dag_{i}d^{}_{i}.
\end{equation}
where $t_{ij}^{\sigma\sigma'}=\exp[\mathrm{i}\mathbf{Q}(\mathbf{R}_i-\mathbf{R}_j)\sigma^x]_{\sigma\sigma'}t_{ij}$, and
\begin{equation}
z_{i\sigma}=(1-d_i^\dag d^{}_i-p^\dag_{i\sigma}p^{}_{i\sigma})^{-1/2}(e^\dag_ip^{}_{i\sigma}+p^\dag_{i\bar\sigma} d^{}_i)\times
\end{equation}
$$
\\\times(1-e_i^\dag e^{}_i-p^\dag_{i\bar\sigma}p^{}_{i\bar\sigma})^{-1/2}.
$$
The Eq. (\ref{eq:H_ext}) is equivalent to (\ref{eq:original_H}) if the unphysical site states (which have no equivalents for the original electronic system) are excluded. The range of definition of $\mathcal{H}_{\rm ext}$ is restricted by the following constraints: 
\begin{equation}
      \label{eq:lmb_constraint}
      d^\dag_id^{}_i+p^\dag_{i\sigma}p^{}_{i\sigma}=c^\dag_{i\sigma}c^{}_{i\sigma},
\end{equation}
\begin{equation}
      \label{eq:eta_constraint}
      e^\dag_ie^{}_i+\sum_\sigma p^\dag_{i\sigma}p^{}_{i\sigma}+d^\dag_id^{}_i=1.
\end{equation}

The presence of the constraints can be taken into account within the functional integral formalism through the Lagrange multipliers (on-site electron energy shift $\lambda_{i\sigma}$ for Eq. (\ref{eq:lmb_constraint}) and local bosonic ``chemical potential'' $\eta_i$ for Eq. (\ref{eq:eta_constraint})) introduced into the action. Up to now the transformation considered is exact, but further exact functional treatment of both bosonic and fermionic fields is hardly possible.
A reasonable  physical picture at large $U/t$ (different from the Hartree--Fock's) can be obtained within the saddle--point approximation for bosonic fields and Lagrange multipliers: for the action $\mathcal{S}$ generated by Eq. (\ref{eq:H_ext}) 
we replace the bosonic fields by their extremal {\it real} values which are assumed to be site-- and time--independent: $e^\dag_i,e^{}_i\rightarrow e;\; p^\dag_{i\sigma},p^{}_{i\sigma}\rightarrow p^{}_{\sigma};\; d^\dag_i,d^{}_i\rightarrow d;\; \eta_i\rightarrow \eta,\; \lambda_{i\sigma}\rightarrow \lambda_{\sigma}$.


The fermionic part of the action can be produced by the Hamiltonian which has the form
\begin{equation}
	\mathcal{H}_{\rm f} = \sum_{\sigma\sigma'ij} z_\sigma z_{\sigma'}(t_{ij}^{\sigma\sigma'}+\delta_{ij}\delta_{\sigma\sigma'}\lambda_\sigma) c^\dag_{i\sigma}c^{}_{j\sigma'},
\end{equation}
which eigenvalues (antiferromagnetic subbands) yield the renormalization of the electronic spectrum:
\begin{equation}
	\label{eq:subband_spectrum}
	E_{s=\pm1}(\kk)=(1/2)\left((z^2_\uparrow+z^2_\downarrow)e^{\rm s}_{\kk}+\lambda_\uparrow+\lambda_\downarrow\right)+ s\sqrt{D_\kk},
\end{equation}
where
\begin{equation}
  D_\kk=(1/4)\left((z^2_\uparrow-z^2_\downarrow)e^{\rm s}_{\kk}+ \lambda_\uparrow-\lambda_\downarrow\right)^2+(e^{\rm a}_{\kk}z_\uparrow z_\downarrow)^2
\end{equation}
and
$e^{\rm s,a}_{\kk} =(1/2)(t_{\kk+\mathbf{Q}/2}\pm t_{\kk-\mathbf{Q}/2})$, $t_\kk = \sum_{i}\exp(\I\kk(\mathbf{R}_i-\mathbf{R}_j))t_{ij}$ is a bare electronic spectrum.

On the thermodynamic level the impact of the on--site states on the electronic states manifests itself in two types of bare spectrum renormalization: narrowing of the bare spectrum  (like in the Hubbard--I approximation\cite{Hubbard}), which is specified by factor $z^2_\sigma<1$  and the additional energy shift $\lambda_\sigma$ (like in the Hartree--Fock approximation where $\lambda^{\rm HF}_{\sigma}=Un/2-Um\sigma,$ $n=\sum_\sigma\langle c^\dag_{i\sigma}c^{}_{i\sigma}\rangle$ is the electronic density and $m=(1/2)\sum_\sigma\sigma \langle c^\dag_{i\sigma}c^{}_{i\sigma}\rangle$ is the site magnetization).
Both $z^2_\sigma$ and $\lambda_\sigma$ are essentially spin--dependent which allows to investigate the formation of magnetic states.
However, in the slave boson formalism, in contrast with HFA, the one--electron energy shift $\lambda_\sigma$ cannot be expressed  in terms of $n$ and $m$ only and should be specified separately, see Eq. (\ref{eq:lmb_eq}).
To write SBA equations, it is convenient to introduce an analogue of the Harris--Lange shift~\cite{Harris-Lange1967}
\begin{equation}
	\label{eq:Phi_def}
	\Phi_{\sigma}\equiv\frac{ep_\sigma+p_{\bar\sigma}d}{(e^2+p_{\bar\sigma}^2)(p_{\sigma}^2+d^2)}\frac1{N}\sum_{\kk s}f(E_s(\kk))\frac{\partial E_s(\kk)}{\partial (z^2_{\sigma})},
\end{equation}
where $f(E)$ is the Fermi function and the derivative $\partial E_s(\kk)/\partial (z^2_{\sigma})$ corresponds to formal derivative of Eq. (\ref{eq:subband_spectrum}) with respect to $z^2_\sigma$.
Then we can write the saddle--point (mean--field) equations:
\begin{eqnarray}
	\label{eq:n_eq}
	n &=& \frac1{N}\sum_{\kk s}f(E_s(\kk)),\\
	\label{eq:m_eq}
	m &=& \frac1{4N}\sum_{\kk s}sf(E_s(\kk))\frac{e^{\rm s}_\kk(z^2_\uparrow-z^2_\downarrow)+\lambda_\uparrow-\lambda_\downarrow}{\sqrt{D_\kk}},\\
	\label{eq:lmb_eq}	
	\lambda_\sigma &=&\nu\left[ \Phi_{\sigma}\left(\frac{p_{\bar\sigma}/e}{e^2+p_{\bar\sigma}^2} + \frac{d/p_{\sigma}}{p_{\sigma}^2+d^2}\right)+\Phi_{\bar\sigma}/(ep_\sigma)\right],
\end{eqnarray}
where we have introduced $\nu = ed-p_\uparrow p_\downarrow$.

The saddle--point values of bosonic variables should be obtained as a solution of self--consistent equations:
\begin{eqnarray}
	\label{eq:n_func}
	2d^2+p_\uparrow^2+p_\downarrow^2&=&n,\\
	\label{eq:m_func}
	(1/2)(p_\uparrow^2-p_\downarrow^2)&=&m,\\
	\label{eq:constraint}
	e^2+p_\uparrow^2+p_\downarrow^2+d^2&=&1,\\
	\label{eq:main_p}
	\nu\sum_\sigma \left(\frac1{ep_{\bar\sigma}}+\frac1{p_\sigma d}\right)\Phi_\sigma &=& U.
\end{eqnarray}

Note that the HFA equations can be easily obtained from the Eqs. (\ref{eq:n_eq},\ref{eq:m_eq},\ref{eq:lmb_eq},\ref{eq:n_func},\ref{eq:m_func},\ref{eq:constraint},\ref{eq:main_p}) by the following ansatz: Eq. (\ref{eq:main_p}) should be replaced by the equation $\nu = 0,\,\lambda_\sigma\rightarrow \lambda^{\rm HF}_\sigma$.
Note that $\nu=0$ implies the vanishing of $\lambda_\sigma$ and these replacements (especially at large $U/t$) violate the Eq. (\ref{eq:main_p}), which indicates insufficiency of HFA. 
The condition $\nu = 0$ is equivalent (by the Schwarz inequality) to $z^2_\uparrow = z^2_\downarrow = 1$, which means that electronic motion is insensitive to the character of site states, even in the sense of averaging single  and double site occupancies ($n_{\sigma}=p^2_\sigma + d^2$).

The saddle point expression for the thermodynamic potential of the spiral state(per site) $\Omega(\mathbf{Q})$ has a form
\begin{equation}
	\label{eq:Omega_sdl_pt}
	\Omega({\bf Q})=Ud^2-\sum_\sigma\lambda_\sigma(p_\sigma^2+d^2)+\Omega_{\rm f}(z^2_\sigma,\lambda_\sigma),
\end{equation}
where $\Omega_f(z^2_\sigma,\lambda_\sigma)$ is the standard potential for the non--interacting electron system in the field $\lambda_\sigma$, with narrowed (by factor $z^2_\sigma$) spectrum
\begin{equation}
	\label{eq:Omega_f_definition}
	\Omega_{\rm f}(z^2_\sigma,\lambda_\sigma) \equiv -(T/N)\sum_{\kk s}\log\left(1+\exp(-E_s(\kk)/T\right) .
\end{equation}

The resulting wave vector is determined by the minimization of $\Omega$ over various spiral states (spiral wave vector $\bf{Q}$).
The minimization of $\Omega(\mathbf{Q})$ with respect to $\bf Q$ was performed numerically while $\textbf{Q}$ was running in the most relevant region of its parameter space; the step of changing its components is $0.01\pi$ (here and hereafter the lattice parameter $a$ is taken to be unity). 
Since $\Omega$  in the ground state actually depends on the chemical potential $\mu$ as a parameter, we can determine the dependence of the magnetic structure on $\mu$  by a procedure taking into account the possibility of the phase separation~\cite{Igoshev10}.

\section{Results}
\label{sec:results}

\subsection{2D lattices}
\label{2D_lattices}
\begin{figure}[h]
\includegraphics[width=0.5\textwidth]{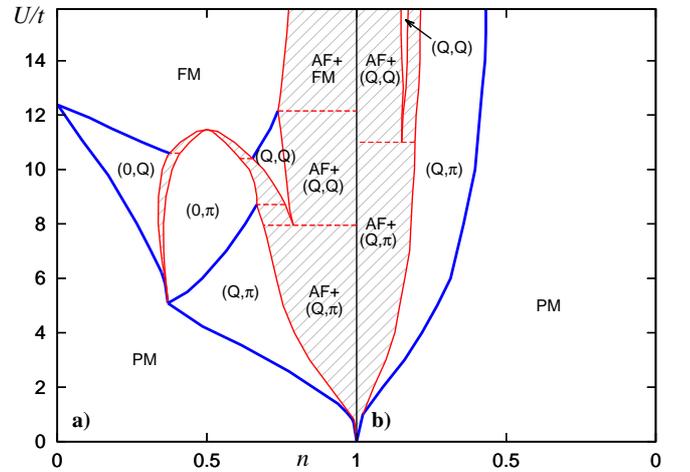}
  \caption{
  Ground state magnetic phase diagram of the Hubbard model for the square lattice with $t'=0$ at $n<1$ within (a) HFA~\cite{Igoshev10} and (b) SBA.
  The phase diagram for the case $n > 1$ due to the electron--hole symmetry ($n\leftrightarrow2-n$) coincides with the given one.
  The spiral phases are denoted according to the form of their wave vector.
  Filling shows the regions of the phase separation.
  Bold lines denote the second order phase transitions.
  Solid lines correspond to the boundaries between the regions of the homogeneous phase and the phase separation.
  Dashed lines show the regions of the separation of different phases.
  $\mathbf{Q}_{\rm AFM} = (\pi,\pi)$
}
\label{fig:square_t'=0}
\end{figure}

We calculated the phase diagrams of the ground state at different $t'/t$ values within the Hartree--Fock approximation in~\cite{Igoshev10}; for $t' = 0$ we present these results in Fig. \ref{fig:square_t'=0}a).
Analogous phase diagram obtained within SBA is shown in Fig. \ref{fig:square_t'=0}b~\cite{Igoshev2013}.
Note that for both these approaches the phase transitions between the different magnetic states are generally first-order transitions that leads to considerable regions of phase separation.
The account of PS considerably distinguishes the phase diagram obtained within SBA from that presented in~\cite{Fresard92}, where PS was disregarded.
The separation regions between the antiferromagnetic phase and the spiral magnetic states (parallel, $\mathbf{Q}=(Q,\pi)$ and diagonal, $\mathbf{Q}=(Q,Q)$) are especially wide.
Therefore, the regions of the pure spiral states are narrowed. In particular, this refers to the diagonal phase, the existence of which becomes possible only at $U > 11t$.
The phase transition between the paramagnetic and spiral magnetic states is of a second order.

One can see that the electron correlations lead to a noticeable suppression of the magnetically ordered states in comparison with HFA: the corresponding concentration intervals in the phase diagram decrease strongly, and the variety of the spiral states disappears.
Besides that, in the slave boson approximation there occurs a wide region of paramagnetic state.
The ferromagnetic state covering a considerable part of the diagram within HFA is shifted to the region of large values $U\gtrsim60t$.
This behavior reproduces the result obtained in~\cite{Fresard92} and is in a good agreement with the variational study of the stability of the saturated ferromagnetic phase~\cite{Linden91}.
The region of the separation of the antiferromagnetic and spiral phases is narrowed by about a factor of 2.

According to our calculations, even  unrestricted increase of $U$ does not make the magnetically ordered states stable far from half filling: at $U=\infty$, there are no spiral magnetic solutions to  equations of the slave boson method at $n < 0.37$ and $n >1.63$.
At the same time, the saturated ferromagnetic solution becomes more favorable than the spiral ones at $1-n < 0.3$.
Thus, the spiral magnetic state at large $U$ values far from half filling replaces the saturated ferromagnetic one.
In contrast to~\cite{Irkhin04}, the unsaturated ferromagnetic solutions exist within our approach, but they are always energy unfavorable in comparison with the saturated ferromagnetic or spiral magnetic states.



\begin{figure}[h]
\includegraphics[width=0.5\textwidth]{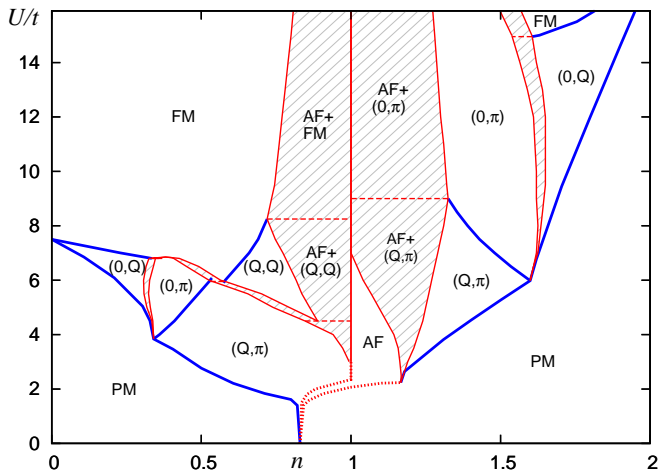}
  \caption{
  Ground state magnetic phase diagram for the square lattice with $t'=0.2$ within HFA~\cite{Igoshev10}.
   Dashed lines denote the first order phase transitions in the case where the region of the phase separation is narrow.
   The notations are analogous to that in Fig.~\ref{fig:square_t'=0}
}
\label{fig:square_t'=0.2_HF}
\end{figure}
\begin{figure}[h]
  \includegraphics[width=0.5\textwidth]{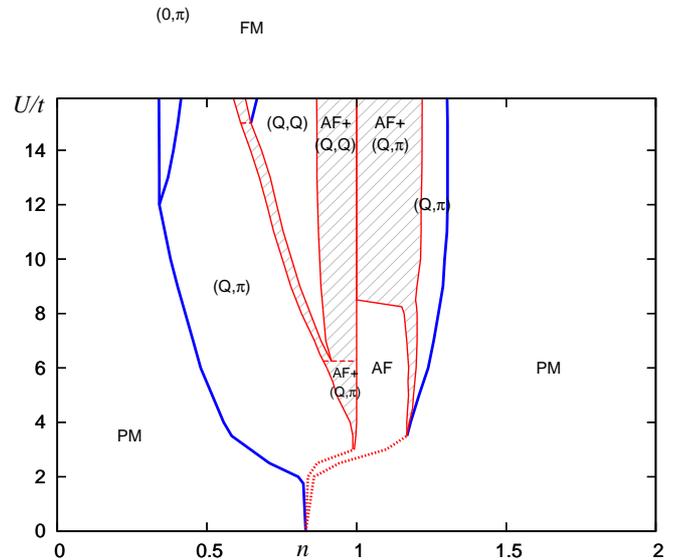}
  \caption{
  The same as in Fig. \ref{fig:square_t'=0.2_HF} within SBA
}
\label{fig:square_t'=0.2_SB}
\end{figure}

Up to the end of this Section we present our results for the square lattice for nonzero value of $t'$ for which the particle--hole symmetry breaks down: The difference of hole--doped ($n<1$) and electron--doped ($n>1$) systems properties is of a great interest\cite{Igoshev10}.
We present the phase diagram for $t'=0.2t$ within HFA (Fig. \ref{fig:square_t'=0.2_HF}) and SBA (Fig. \ref{fig:square_t'=0.2_SB}).
In comparison with the case $t'=0$, for the case $n<1$ the diagonal phase shifts to the region of much smaller $U/t$ values and the parallel phase becomes more extended over the concentration parameter.
The physical origin is the shift of the density value $\nVH$ corresponding to the coincidence of the Fermi level in the paramagnetic phase and the van Hove singularity level (for $t'=0.2t$ $n_{\rm vH}\approx0.83$) with respect to half filling ($n = 1$): at $t'\ne0$ these points give their own tendencies to the corresponding magnetic ordering.
We see that in the hole--doped half of the phase diagram the correlation effects lead only to a quantitative renormalization of the phase boundaries.

In the case of $n > 1$, the correlation effects are more significant: all homogeneous spiral states disappear, except for a narrow region of the parallel phase.
Such a suppression of magnetism is explained by the fact that here all the singular features  come from the point  $n = 1$.
Far from half filling, any spiral magnetism is impossible and the saturated ferromagnetism is unfavorable in comparison with the paramagnetic phase at any large $U$ value, when correlations are taken into account.
Thus, the electron--hole asymmetry is enhanced considerably in comparison with HFA.

These results can be used for qualitatively explaining of the magnetic properties of layered high--temperature superconducting perovskites, for which the fitting of the angular resolved photoemission spectroscopy spectra to the bare model one yields $t'\sim 0.2t$.
The results are in agreement\cite{Igoshev2013} with the experimental data on the magnetic structure of the hole--doped compound La$_{2-p}$Sr$_p$CuO$_4$, which has a close value of the $t'/t$ parameter~\cite{Tanamoto93,Matsuda02,Fujita02,Yamada98}.
At the same time, for the high temperature superconducting compound Nd$_{2-x}$Ce$_x$CuO$_4$, in which the charge carriers are electrons, the homogeneous commensurate antiferromagnetic ordering is stable up to $x = 0.14$~\cite{Takagi89} in agreement with our results for $n > 1$ (see Fig. \ref{fig:square_t'=0.2_SB}).

\subsection{3D lattices}
\label{3D_lattices}

\begin{figure}[h]
\includegraphics[width=0.5\textwidth]{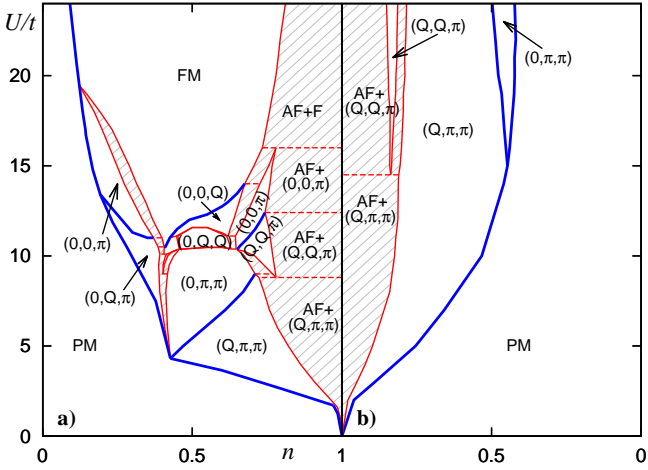}
\caption{
	Ground state magnetic phase diagram for the simple cubic lattice with $t'=0$ at $n<1$ within (a) HFA and (b) SBA.
	The notations are analogous to that in Fig.~\ref{fig:square_t'=0}.
  At second order transitions between collinear AFM and spiral phases the wave vector $\mathbf{Q}$ reaches continuously the Brillouin zone boundary ($Q = \pi$);
  $\mathbf{Q}_{\rm AFM} = (\pi,\pi,\pi)$
}
\label{fig:cubic_t'=0}
\end{figure}

Now we present  the results for three--dimensional lattices: simple cubic (sc) and body centered cubic (bcc) lattices.
First we consider the ground state magnetic phase diagram for sc lattice with $t'=0$.
The physical picture is very similar to that for the square lattice (see Fig. \ref{fig:cubic_t'=0} for  comparison of the results of HFA (a) and SBA (b) approaches): the density value $n=1$ corresponding to both van Hove singularity and perfect antiferromagnetic nesting peculiarity retains its crucial role.
The spiral $\textbf{Q} = (Q,\pi)$ magnetic phase for the square lattice is replaced by the spiral with ${\bf Q}=(Q,\pi,\pi)$.
Note that finiteness of the density of states at the Fermi level corresponding to $n=1$ in the paramagnetic phase (which makes the tendency to ferromagnetic ordering weaker within the Stoner theory) does not actually affect the phase diagram since this tendency is  in any case much weaker than the antiferromagnetic instability tendency present in the vicinity of half--filling.

\begin{figure}[h]
\includegraphics[width=0.5\textwidth]{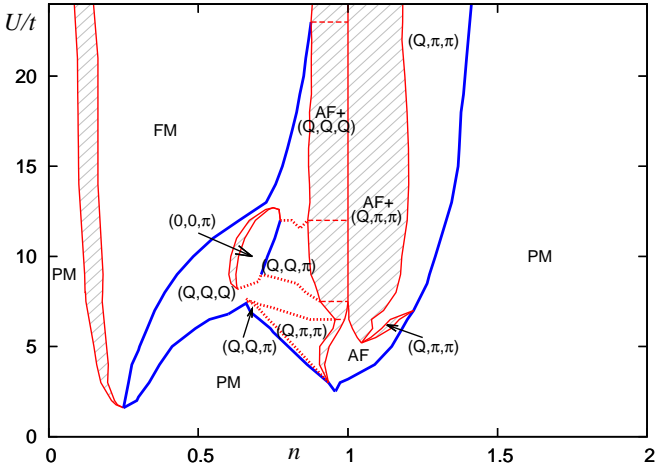}
\caption{
Ground state magnetic phase diagram for the simple cubic lattice with $t'=0.3$ within SBA.
  $\mathbf{Q}_{\rm AFM} = (\pi,\pi,\pi)$
}
\label{fig:cubic_t'=0.3_SB}
\end{figure}

We consider the results of SBA in the case of non--zero $t'$ (see Fig. \ref{fig:cubic_t'=0.3_SB}) where these tendencies are well separated.
One can see that in the vicinity of $\nVH\approx 0.25$ and rather small $U/t\approx 1.5$ the competition of diagonal ($\mathbf{Q}=(Q,Q,Q)$) and ferromagnetic phases is present to the right of $n_{\rm vH}$ (this has a direct analogy in the square lattice case).
On the other hand, there is PS into the paramagnetic and ferromagnetic phases to the left of point $\nVH$; this does not occur in the two--dimensional case.
However, the ferromagnetic region is rather narrow and survives (as well as the diagonal phase) in the vicinity of $\nVH$; this feature is  different from the square lattice.
As well as in the 2D case, we found hole--particle strong asymmetry: for $n<1$ the variety of magnetic phases appears at not too large $U/t$.
For instance, at a realistic value $U=4t$ we found AFM phase in the vicinity of half--filling, which is replaced by the phase with $\mathbf{Q}=(Q,\pi,\pi)$ (and a small fracture of $(Q,Q,\pi)$ phase) as the density becomes lower. Then one observes a wide non--magnetic region, and at low density the system enters the region of ferromagnetic and diagonal phase competition.
For $n>1$ the magnetism is strongly suppressed (as well as in 2D case).

\begin{figure}[h]
\includegraphics[width=0.5\textwidth]{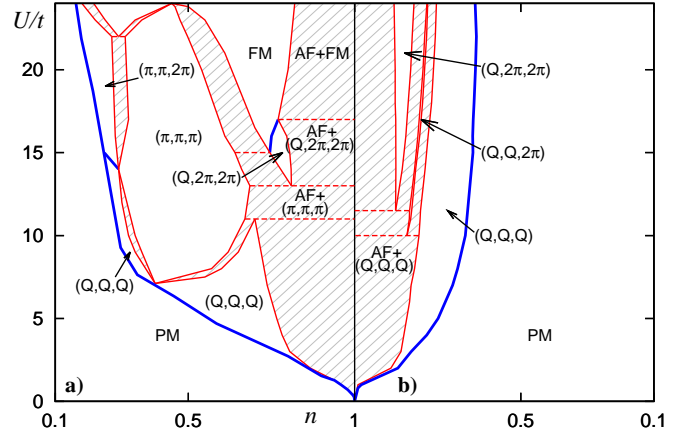}
\caption{
Ground state magnetic phase diagram for the body centered cubic lattice with $t'=0$ at $n<1$ within (a) HFA and (b) SBA.
 At second order transitions  the wave vector $\mathbf{Q}$ reaches continuously the Brillouin zone boundary ($Q = \pi$ or $2 \pi$);
  $\mathbf{Q}_{\rm AFM} = (0,0,2\pi)$
}
\label{fig:bcc_t'=0}
\end{figure}

In Fig. \ref{fig:bcc_t'=0} we present the results for the bcc lattice Hubbard model in the case $t'=0$.
One can see that correlation effects strongly but only quantitatively renormalize the boundary phase lines in a rather wide vicinity of half--filling, but well away from half--filling they fully destroy magnetic ordering.
The van Hove singularity corresponding to $n = \nVH = 1$ results in this case in an especially wide PS and magnetic phase regions; the dominating magnetic order is diagonal one ($\mathbf{Q}=(Q,Q,Q)$).
This is a manifestation of especially strong van Hove singularity at the band centre for the  bcc lattice.

\begin{figure}[h]
\includegraphics[width=0.5\textwidth]{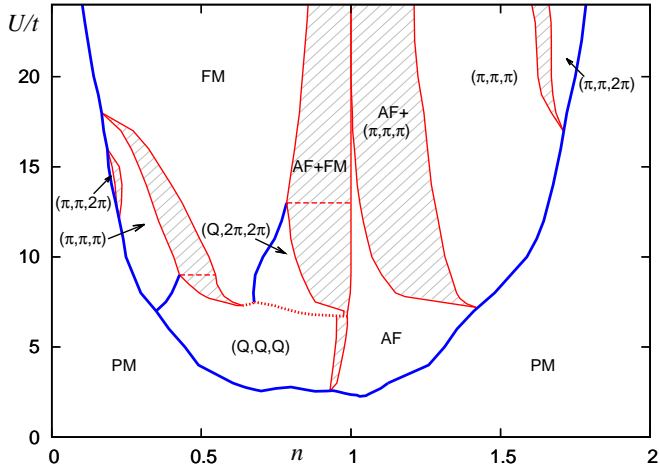}
\caption{
	Ground state magnetic phase diagram for bcc lattice with $t'=0.3$ within HFA.
   $\mathbf{Q}_{\rm AFM} = (0,0,2\pi)$
}
\label{fig:bcc_t'=0.3_HF}
\end{figure}

\begin{figure}[h]
\includegraphics[width=0.5\textwidth]{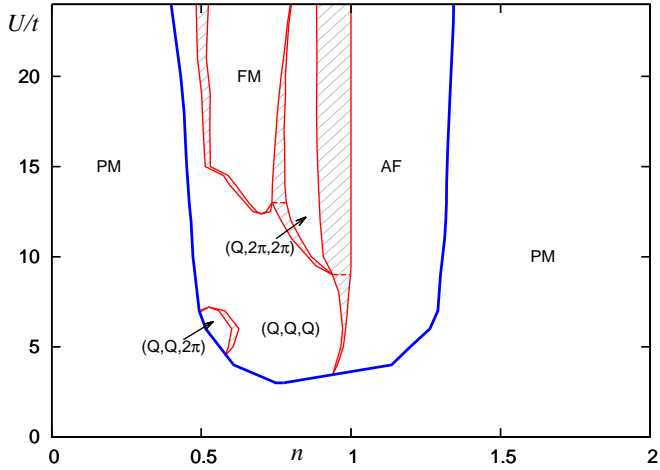}
\caption{
	Ground state magnetic phase diagram for bcc lattice with $t'=0.3$ within SBA.
  $\mathbf{Q}_{\rm AFM} = (0,0,2\pi)$
}
\label{fig:bcc_t'=0.3_SB}
\end{figure}

We present the results for bcc lattice with $t'=0.3t$ obtained within HFA (Fig. \ref{fig:bcc_t'=0.3_HF}) and SBA (Fig. \ref{fig:bcc_t'=0.3_SB}).
It is evident that the results are considerably different from square and sc lattices.
For electron doping the correlation effects results in the fact that up to large $U/t$ pure antiferromagnetic ordering in the vicinity of half--filling occurs, which is changed by paramagnetic phase for $n\gtrsim 1.4$. The variety of magnetic phases which exist far away from half--filling in HFA vanishes.
For hole doping at rather large $U/t\gtrsim 10$  there occur  ferromagnetic phase and $\mathbf{Q}=(0,0,Q)$ phase which can be considered as a modulated ferromagnetic phase.
At smaller $U/t$ the  diagonal $(0,0,Q)$ phase becomes favorable in a  wide region. In HFA there exists PS into ferromagnetic and AFM phases in the vicinity of half--filling.
It is interesting that in this case  correlations practically do not renormalize critical values of $U$. However the structure of PS regions is strongly changed.

\section{Conclusions and Discussion}
\label{sec:conclusions}

The results obtained can be applied to analysis of properties and phenomena in $d$--metals and their compounds, especially with fcc and bcc structures.
The correlations effects considered can be important, e.g.~for metal--insulator (Mott) transitions.
Within the Hartree--Fock approximation including spiral states the latter problem was treated in~\cite{Timirgazin12}.

The slave boson approximation enables one to take into account correlations in terms of a few parameters renormalizing the electron spectrum.
Such an approach is more simple and transparent in comparison with dynamical mean--field theory (DMFT) which employs local approximation for the self--energy, so that spectrum renormalization is difficult for the interpretation.
We see that correlation effects lead to  strong suppression of the regions of the existence of the magnetic phases.
At the same time, the first-order transitions and the noticeable regions of the phase separation between the magnetically ordered states are retained.
The correlation effects near half filling change only slightly  the Hartree--Fock results, so that at small  $t'/t$ they do not modify the sequences of the magnetic states with increasing $U$.

The increase in the $t'/t$ parameter leads to a redistribution of the electron density of states closer to the band bottom and to the van Hove singularity, which is important for  formation of magnetism.
Thus the correlations result in a considerable modification of the phase diagram. In particular, the paramagnetic region occurs in the SB  method at large $U$.
Besides that, the asymmetry of the magnetic phases  with respect to the sign of  charge carriers increases in comparison with the Hartree--Fock approximation.
When the concentration is far from half filling and the Fermi level is far from the van Hove singularity, the magnetic state cannot be formed at any $U$.

\section{Acknowledgements}
\label{sec:akhnoledgments}
This work was supported in part by the Division of Physical Sciences and Ural Branch of RAS (project nos. 12--T--2--1001,  14--2--NP-273); by the Presidium of RAS (project nos. 12--P--2--1041, 12--U--2--1021); and by the Russian Foundation for Basic Research (project nos. 12--02--00632-a, 14--02--31603--mol--a). Most calculations were performed using ``Uran'' cluster of IMM of RAS.




\begin{thebibliography}{60}


\bibitem{Nagaoka66} Y.~Nagaoka, Phys. Rev. \textbf{147}, 392 (1966).


\bibitem{Linden91} W.~von~der~Linden and D.\,M. Edwards, J. Phys.: Cond. Matt. \textbf{3}, 4917 (1991);
P.~Wurth, G.S.~Uhrig and E.~M\"uller--Hartmann, Ann. Phys. \textbf{2}, 13960 (1997).

\bibitem{Irkhin04} V.Yu.~Irkhin and A.V.~Zarubin, Phys. Rev. B \textbf{70}, 035116 (2004).
\bibitem{Visscher73} P.B.~Visscher, Phys. Rev. B \textbf{10}, 943 (1973).
\bibitem{Emery90} V.J.~Emery, S.A.~Kivelson and H.Q.~Lin, Phys. Rev. Lett. \textbf{64}, 475 (1973).
\bibitem{Schultz90} H.J.~Schultz, Phys. Rev. Lett. \textbf{64}, 1445 (1990).

\bibitem{Arrigoni91} E.~Arrigoni and G.C.~Strinati, Phys. Rev. B \textbf{44}, 7455 (1991).
\bibitem{Igoshev10} P.A.~Igoshev, M.A.~Timirgazin, A.A.~Katanin, A.K.~Arzhnikov and V.Yu.~Irkhin, Phys. Rev. B \textbf{81},
094407 (2010).

\bibitem{Shraiman90} B.~Shraiman and E.~Siggia, Phys. Rev. Lett. \textbf{62}, 1564 (1990).



\bibitem{HTSCrewiew} M.A.~Kastner, R.J.~Birgeneau, G.~Shirane and Y.~Endoh, Rev. Mod. Phys.  \textbf{70}, 897 (1998).


\bibitem{Bao09} W.~Bao, Y.~Qiu, Q.~Huang et al., Phys. Rev. Lett. \textbf{102}, 247001 (2009);
D.K.~Pratt, M.G.~Kim, A.~Kreyssig et al., Phys. Rev. Lett. \textbf{106}, 257001 (2011);
Z.~Xu, J.~Wen, Y.~Zhao et al., Phys. Rev. Lett. \textbf{109}, 227002 (2012);
J.W.~Lynn and P.~Dai, Physica C \textbf{469}, 469 (2009).

\bibitem{Ru} A.P.~Mackenzie and Y.~Maeno, Rev. Mod. Phys. \textbf{75}, 657 (2003).

\bibitem{Igoshev11} P.A.~Igoshev, V.Yu.~Irkhin, and A.A.~Katanin, Phys. Rev. B \textbf{83}, 245118 (2011).

\bibitem{Igoshev07} P.A.~Igoshev, A.A.~Katanin, V.Yu.~Irkhin, JETP \textbf{105}, 1043 (2007).

\bibitem{Sarker91} S.~Sarker, C.~Jayaprakash, H.R.~Krishnamurthy and W.~Wenzel, Phys. Rev B \textbf{43}, 8775 (1991).

\bibitem{Kivelson03} S.A.~Kivelson,  I.P.~Bindloss, E.~Fradkin et al.,  Rev. Mod. Phys. \textbf{75}, 1201 (2003).

\bibitem{Vedyaev} M.A.~Timirgazin, A.K.~Arzhnikov and A.V.~Vedyayev, Sol. St. Phen. \textbf{190}, 67 (2012).

\bibitem{Kotliar86} G.~Kotliar and A.E.~Ruckenstein, Phys. Rev. Lett. \textbf{57}, 1362 (1986).

\bibitem{Gutzwiller} M.C.~Gutzwiller, Phys. Rev. \textbf{137}, A1726 (1965).


\bibitem{Fresard92} R.~Fresard and P.~W\"{o}lfle, J. Phys.: Cond. Matt. \textbf{4}, 3625 (1992).

\bibitem{Fresard98} R.~Fresard and W.~Zimmermann, Phys. Rev. B \textbf{58}, 15288 (1998).

\bibitem{Moriya} T.~Moriya, \textit{Spin Fluctuations in Itinerant Electron Magnetism}, Springer: Berlin, 1985.

\bibitem{Hubbard2} J.~Hubbard, Proc. Roy. Soc. Series A. \textbf{277}, 237 (1964).

\bibitem{Hubbard} J.~Hubbard, Proc. Roy. Soc. Series A. \textbf{276}, 238 (1963).

\bibitem{Harris-Lange1967} A.B.~Harris and R.V. Lange, Phys. Rev. B \textbf{157},295 (1967).

\bibitem{Igoshev2013} P.A.~Igoshev, M.A.~Timirgazin, A.K.~Arzhnikov and V.Yu.~Irkhin, JETP Letters \textbf{98}, 150 (2013).


\bibitem{Fujita02} M.~Fujita, K.~Yamada, H.~Hiraka et al., Phys. Rev. B \textbf{65}, 064505 (2002).

\bibitem{Tanamoto93} T.~Tanamoto, H.~Kohno and H.~Fukuyama, J. Phys. Soc. Jpn. \textbf{62}, 717 (1993).

\bibitem{Matsuda02} M.~Matsuda, M.~Fujita, K.~Yamada et al., Phys. Rev. B \textbf{65}, 134515 (2002).

\bibitem{Yamada98} K.~Yamada, C.H.~Lee, K.~Kurahashi et al., Phys. Rev. B \textbf{57}, 6165 (1998).

\bibitem{Takagi89} H.~Takagi, S.~Uchida, and Y. Tokura, Phys. Rev. Lett. \textbf{62}, 1197 (1989).




\bibitem{Timirgazin12} M.A.~Timirgazin, A.K.~Arzhnikov and V.Yu. Irkhin, JETP Letters \textbf{86}, 171 (2012).









\end{thebibliography}


\end{document}